\documentclass[prd, preprint, nofootinbib, superscriptaddress]{revtex4}
\usepackage{epsfig,graphicx,amsmath,amssymb}
\usepackage{graphicx}
\usepackage{amsfonts}
\usepackage{latexsym}
\usepackage{amssymb}
\usepackage{slashed}
\usepackage{subfigure} 
\usepackage{here}
\usepackage{ulem}

\usepackage[bookmarks=true, bookmarksnumbered=true,bookmarkstype=toc, colorlinks=true, 
citecolor=blue, linkcolor=blue, urlcolor=blue]{hyperref} 
\usepackage{units}

\usepackage[dvipsnames]{xcolor}

\preprint{UME-PP-027}
\preprint{EPHOU-24-003} 

\begin{document}

\title{Freeze-in sterile neutrino dark matter \\ in a feebly gauged $B-L$ model}

\author{Osamu Seto}
 \email{seto@particle.sci.hokudai.ac.jp}
\affiliation{Department of Physics, Hokkaido University, Sapporo 060-0810, Japan}

\author{Takashi Shimomura}
\email{shimomura@cc.miyazaki-u.ac.jp}
\affiliation{Faculty of Education, Miyazaki University, Miyazaki 889-2192, Japan}

\author{Yoshiki Uchida}
\email{uchida.yoshiki@m.scnu.edu.cn}
\affiliation{State Key Laboratory of Nuclear Physics and Technology, Institute of Quantum Matter, South China Normal University, Guangzhou 510006, China}
 \affiliation{Guangdong Basic Research Center of Excellence for Structure and Fundamental Interactions of Matter, Guangdong Provincial Key Laboratory of Nuclear Science, Guangzhou 510006, China}

%
\begin{abstract}
We consider the gauged $U(1)_{B-L}$ model and examine the situation where the sterile neutrino is a dark matter candidate produced by the freeze-in mechanism. 
In our model, the dark matter $N$ is mainly produced by the decay of a $U(1)_{B-L}$ breaking scalar boson $\phi$.
We point out that the on-shell production of $\phi$ through annihilation of the $U(1)_{B-L}$ gauge boson $Z'$ plays an important role. 
We find that the single production of $Z’$ from the gluon bath in the early Universe can become the main production modes for $Z'$ in some parameter regions.
To prevent $N$ from being overproduced, we show that the $U(1)_{B-L}$ gauge coupling constant $g_{B-L}$ must be as small as $10^{-16}$--$10^{-10}$. 
We also consider the case where the decay of $\phi$ into $N$ is kinematically forbidden. 
In this case, $N$ is generated by the scattering of $Z'$ and the $g_{B-L}$  takes values of $10^{-10}$--$10^{-6}$, which can be explored in collider experiments like FASER and SHiP.

\end{abstract}

\vspace*{1cm}
\maketitle


\section{Introduction}

Nonvanishing masses of neutrinos indicate the existence of the standard model (SM) gauge singlet right-handed (RH) neutrinos.
Such a RH neutrino would be referred to as sterile neutrino because those do not interact 
through the weak interaction, except through possible tiny mixing with left-handed (LH) active neutrinos. 
Massive sterile neutrinos are electrically neutral and could be very long lived if the mixing with active neutrinos, in other words Yukawa couplings, is small enough. Then one of the RH neutrinos is a good candidate for dark matter (DM) in our Universe~\cite{Dodelson:1993je,Dolgov:2000ew,Asaka:2005an,Asaka:2005pn} (for a review see e.g., Refs.~\cite{Drewes:2016upu,Boyarsky:2018tvu}), while the others can generate neutrino masses through so-called seesaw mechanism~\cite{Minkowski:1977sc,Yanagida:1979as,GellMann:1980vs,Mohapatra:1979ia}.

In the minimal scenario of sterile neutrino DM $N$, sterile neutrinos never reach thermal equilibrium but are thermally produced as warm DM, 
by so-called Dodelson-Widrow mechanism, through the mixing with active neutrinos~\cite{Dodelson:1993je}. However such sterile neutrino DM 
have been confronted with astrophysical x-ray bounds and Lyman-$\alpha$ constraints on its free streaming scale~\cite{Boyarsky:2005us,Boyarsky:2006fg,Boyarsky:2006ag,Boyarsky:2007ay,Yuksel:2007xh}\footnote{The production by the mixing is available in the case that the production rate is enhanced by the resonance effect under a large lepton asymmetric background~\cite{Shi:1998km}. See, for a recent study on cosmological constraints on large lepton asymmetry, e.g., Ref.~\cite{Seto:2021tad}.}. 
We may seek another production mechanism in an extended model of the SM where an additional interaction besides the mixing exists.
While RH neutrinos are singlet under the SM, they might be charged under an additional gauge symmetry.
Gauged $B-L$ (Baryon number $-$ Lepton number) symmetry is a theoretically well motivated extension of the SM~\cite{Davidson:1978pm,Mohapatra:1980qe,Marshak:1979fm}.
Since a RH neutrino carries lepton number, sterile neutrinos are charged under the $U(1)_{B-L}$.
Production of sterile neutrinos through the $U(1)_{B-L}$ gauge boson $Z'$ is a viable alternative~\cite{Khalil:2008kp} in the minimal $U(1)_{B-L}$ model where one scalar to break the $B-L$ gauge symmetry and three RH neutrinos to cancel all anomalies are minimally introduced. 
Then, if the $B-L$ gauge coupling constant $g_{B-L}$ is of the order of unity, sterile neutrinos would be easily thermalized by the new $U(1)_{B-L}$ gauge interaction, which would give rise to overabundant or too warm, unless the cosmic thermal history is non-standard~\cite{Khalil:2008kp} or parity-odd sterile neutrino~\cite{Okada:2010wd} is as heavy as weakly interacting massive particle~\cite{Okada:2016gsh}. 

On the other hand, for the very small $g_{B-L}$, sterile neutrino DM is generated non-thermally by the scattering or decay in the thermal plasma. Such non-thermal production had been applied for various particles~\cite{Ellis:1984eq,McDonald:2001vt}, which was coined as ``freeze-in'' production~\cite{Hall:2009bx}.  
Processes involving $Z'$~\cite{Kaneta:2016vkq,Biswas:2016bfo,Heeba:2019jho,Seto:2020udg,Okada:2020cue,Okada:2021nwo,Iwamoto:2021fup} appear to be enough to explain the production of sterile neutrino DM. 
However, the $U(1)_{B-L}$ breaking scalar $\phi$ contributions cannot be ignored for theoretical consistency to describe the broken gauge theory.
Moreover, these contributions were shown to be important for the final dark matter abundance  
by solving the simultaneous Boltzmann equations of $N$ and $Z'$
~\cite{Eijima:2022dec}, because the production processes occur most efficiently at a low temperature.

In this paper, to take the on-shell scalar contributions into account for $m_{Z'} < 2m_N$, we solve the set of Boltzmann equations of 
not only sterile neutrino and $Z'$ but also $\phi$ with and without the mixing with the SM Higgs boson.
As the decay and its inverse decay of $Z'$ are dominant processes for production or thermalization of $Z'$~\cite{Eijima:2022dec}, we point out that this is the case for $\phi$ as well. 
 We also point out that the single $Z'$ production mode via gluon is significant in the $Z'$ production.
We find that the required $g_{B-L}$ must be smaller than $\mathcal{O}(10^{-10})$ to reproduce the observed DM abundance for $m_{\phi} > 2m_N$.
On the other hand, for $m_{\phi} < 2m_N$, the cosmologically interesting parameter region, $g_{B-L} \lesssim 10^{-6}$, can be probed by long lived $Z'$ searches as have been reported in the literature~\cite{Kaneta:2016vkq,Okada:2020cue,Eijima:2022dec,Asai:2022zxw}.

This paper is organized as follows. We give the Lagrangian of our model and the decay rate of the singlet-like scalar $\phi$ in Sec.~\ref{sec:model}, and summarize Boltzmann equations to be solved in Sec.~\ref{sec:bolt}.
In Sec.~\ref{sec:abundance}, we show our results for $m_{\phi} > 2m_N$ with emphasizing the contribution of $\phi$ to the DM abundance.
For this parameter range to reproduce the correct DM abundance, all of $N$, $Z'$ and $\phi$ are feeble particles
 and never reach thermal equilibrium. We solve the evolution of the number densities of such multiple spices in the feeble $U(1)_{B-L}$ sector. 
We also discuss cosmological, astrophysical and experimental constraints and implications for the opposite mass spectrum case of $m_{\phi} < 2m_N$.
We conclude this paper in Sec.~\ref{sec:concl}.

\section{Model}
\label{sec:model}

We consider the minimal gauged $U(1)_{B-L}$ model studied in Ref.~\cite{Eijima:2022dec}. 
In this model, three generations of RH neutrinos ($\nu_R^i$,  $i=1, 2, 3$) are introduced to 
cancel the gauge anomalies. One complex scalar $\Phi$, which spontaneously breaks the $B-L$ gauge symmetry, is also introduced as the origin of Majorana mass of RH neutrinos.
The gauge charge assignment is shown in Table~\ref{tab:matter-contents}. Here, $Q$ $(u_R,d_R)$ and $L$ $(e_R,\nu_R)$ denote 
the LH (RH) quarks and leptons, respectively, and the SM Higgs doublet is denoted by $H$.
%
\begin{table}[tb]
	\centering
	\begin{tabular}{|c|ccc|ccc|cc|} \hline	
		                   & ~~$Q^i$~~ & $~~u^i~~$  & $~~d^i~~$ & $~~L^i~~$ & $~~e_R^i~~$ & $~~\nu_R^i~~$ & $~~H~~$ & $~~\Phi~~$ \\ \hline
		$SU(3)_C$ & $\mathbf{3}$ & $\mathbf{3}$  & $\mathbf{3}$ & $\mathbf{1}$ & $\mathbf{1}$ & $\mathbf{1}$ & $\mathbf{1}$ & $\mathbf{1}$\\ 
		$SU(2)_L$ & $\mathbf{2}$ & $\mathbf{1}$ & $\mathbf{1}$ & $\mathbf{2}$ & $\mathbf{1}$ & $\mathbf{1}$ & $\mathbf{2}$ & $\mathbf{1}$\\
		$U(1)_Y$ & $\frac{1}{6}$ & $\frac{2}{3}$ & $-\frac{1}{3}$ & -$\frac{1}{2}$ & $-1$ & $0$ & $\frac{1}{2}$ & $0$ \\ \hline
		$~~U(1)_{B-L}~~$ & $\frac{1}{3}$ & $\frac{1}{3}$ & $\frac{1}{3}$ & $-1$ & $-1$ & $-1$ & $0$ & $2$ \\ \hline
	\end{tabular}
\caption{Particle content and gauge charge assignment. The RH neutrinos are denoted as $\nu_R^i$ ($i=1, 2, 3$) and 
new scalar boson is denoted as $\Phi$.
}
\label{tab:matter-contents}
\end{table}

\subsection{Lagrangian}
The Lagrangian of the model is written as
\begin{align}
 \mathcal{L} &= \mathcal{L}_\text{SM} + \mathcal{L}_{\nu_R} + \mathcal{L}_\text{gauge} + |D_\mu \Phi|^2  - V, \label{eq:full-lag}
\end{align}
where $\mathcal{L}_\text{SM}$ represents the SM Lagrangian without a scalar potential. The Lagrangian regarding the RH neutrino $\mathcal{L}_{\nu_R}$ 
is given by
\begin{align}
 \label{Lag1} 
 \mathcal{L}_{\nu_R} &= \frac{1}{2} \overline{\nu_R^i} (i \slashed{D}) \nu_R^i - {y_{\nu}}_{ij}\overline{L^{i}} \tilde{H} \nu_R^j - \frac{1}{2} {y_{\nu_R}}_i \Phi \overline{\nu_R^{i~C}} \nu_R^i  + \mathrm{ H.c.} ,
\end{align}
where the superscript $C$ denotes the charge conjugation, and $\tilde{H} \equiv \epsilon H^{\dagger}$ is the complex conjugate Higgs field 
with $\epsilon$ being the antisymmetric tensor of $SU(2)_L$.
The Yukawa couplings for the LH lepton doublets and RH neutrinos are denoted by ${y_{\nu}}_{ij}$ and ${y_{\nu_R}}_i$, respectively, in which $i$ 
and $j$ are indices of flavor or generation. Without loss of generality, we can work on the diagonal basis of ${y_{\nu_R}}_i$. 

The covariant derivative in this model is given as 
\begin{align}
D_\mu = \partial_\mu -i g_3 G_\mu - i g_2 W_\mu - i g_1 Y B_\mu - i g_{B-L} Q_{B-L} X_\mu ,
\end{align}
where $G_\mu,~W_\mu$ and $B_\mu$ represent the gauge fields of $SU(3)_C,~SU(2)_L$ and $U(1)_Y$, and 
$g_3,~g_2$ and $g_1$ are the corresponding gauge coupling constants, respectively.  The remaining one $X_\mu$ 
is the gauge boson of $U(1)_{B-L}$, and $g_{B-L}$ is its gauge coupling constant.
The $U(1)_Y$ and $U(1)_{B-L}$ charges are denoted as $Y$ and $Q_{B-L}$, which are shown in Table~\ref{tab:matter-contents}.

The Lagrangian for the $B-L$ gauge boson kinetic term, $\mathcal{L}_\text{gauge}$, consists of 
\begin{align} 
\mathcal{L}_\text{gauge} = -\frac{1}{4} X_{\mu \nu} X^{\mu\nu} + \frac{\varepsilon}{2} B_{\mu\nu} X^{\mu\nu},
\end{align}
where $X_{\mu\nu}$ and $B_{\mu\nu}$ represent the gauge field strength of $X$ and $B$, respectively.  
The second term is the gauge kinetic mixing term with a mixing parameter $\varepsilon$. For the sake of minimality, 
we set the kinetic mixing parameter to be negligibly small and omit this term throughout this paper\footnote{The kinetic mixing can be generated through loop processes. Such a loop-induced kinetic mixing is not finite and should be renormalized. We assume that our kinetic mixing parameter is tiny after the loop induced ones are renormalized.}.

The fourth term of Eq.~\eqref{eq:full-lag} is the kinetic term of $\Phi$, and the last one $V$ is the scalar potential which is given by
\begin{align}
V & = \frac{\lambda_1}{2} \left( |H|^2-\frac{v^2}{2}\right)^2+\frac{\lambda_2}{2} \left( |\Phi|^2-\frac{v_{B-L}^2}{2}\right)^2 
  +\lambda_3 \left(|H|^2-\frac{v^2}{2}\right) \left( |\Phi|^2-\frac{v_{B-L}^2}{2}\right) ,
\label{eq:V}
\end{align}
where $\lambda_1, \lambda_2, \lambda_3$ are real and positive parameters. On the electroweak and $B-L$ broken vacuum, the scalar bosons are expanded around its vacuum expectation value (vev), $v$ and $v_{B-L}$, respectively as
\begin{align}
H= 
\begin{pmatrix}
0 \\
\frac{1}{\sqrt{2}}(v + \phi_H)
\end{pmatrix},~~~~
\Phi = \frac{1}{\sqrt{2}} ( v_{B-L} + \phi_{B-L}),
\end{align}
where $\phi_H$ and $\phi_{B-L}$ are the dynamical degree of freedom. 
Here, the Nambu-Goldstone bosons to be absorbed by the gauge bosons are omitted.

After the spontaneous symmetry breaking of $SU(2)_L \times U(1)_Y $ and $U(1)_{B-L}$ symmetries, the gauge bosons and RH neutrinos acquire the masses.
The neutrinos also acquire the masses via the type-I seesaw mechanism.  
We denote the mass eigenstates of the active, sterile neutrinos as $\nu, N_i$, and also those of the extra gauge and scalar bosons as $Z'$ and $h,~\phi$, respectively. The masses of $N_i$ and $Z'$ are given by
\begin{align}
m_{N_i} &\simeq \frac{{y_{\nu_R}}_i }{\sqrt{2}} v_{B-L}, \label{eq:N-mass} \\
m_{Z'} &= 2 g_{B-L} v_{B-L}, \label{eq:Z'-mass}
\end{align}
where we assume that $v_{B-L}$ is much larger than $v$.  
Using Eqs.~\eqref{eq:N-mass} and \eqref{eq:Z'-mass}, the Majorana Yukawa coupling and vev are expressed by the $N_i$ and $Z'$ mass, respectively, in the following section.  
The masses of $h$ and $\phi$ are given by
\begin{align}
m_h^2 &= \frac{1}{2}\left( \lambda_1 v^2+\lambda_2 v_{B-L}^2 + \frac{\lambda_1 v^2-\lambda_2 v_{B-L}^2}{\cos(2 \alpha )} \right) , \\
m_{\phi}^2 &= \frac{1}{2}\left( \lambda_1 v^2+\lambda_2 v_{B-L}^2 - \frac{\lambda_1 v^2-\lambda_2 v_{B-L}^2}{\cos(2 \alpha )} \right), \label{eq:phi-mass}
\end{align}
where $\alpha$ is the scalar mixing angle defined by
\begin{align}
\label{eq:sinalpha}
\sin (2 \alpha ) & \simeq   \frac{v m_{Z'}}{m_\phi^2 - m_h^2}\frac{\lambda_3}{g_{B-L}}.
\end{align}
The mass eigenstates, $h$ and $\phi$, are related with $\phi_H$ and $\phi_{B-L}$ as
\begin{align}
&
\left( 
\begin{array}{c} 
\phi_H \\
\phi_{B-L} \\
\end{array}
\right)
 =
\left(
\begin{array}{cc}
 \cos\alpha & \sin\alpha  \\
 -\sin\alpha & \cos\alpha \\
\end{array}
\right)
\left( 
\begin{array}{c} 
h \\
\phi \\
\end{array}
\right).
\end{align}
For the rest of this paper, we identify $h$ as the SM Higgs boson with the mass of $125$ GeV.

\subsection{Decay rates of $\phi$}
In our previous work \cite{Eijima:2022dec}, the freeze-in productions of $N$ from nonthermal decays and/or scatterings of $Z'$ were studied by solving the Boltzmann equations. 
Here and hereafter, the DM is denoted as $N$, which is sequestered from the SM particles, one among the three sterile neutrinos. The other $N_i$ responsible to generate neutrino mass are denoted by $\nu_R$.
The relevant decay rates and cross sections of $N$ and $Z'$ can be found in Ref.~\cite{Eijima:2022dec}.  
In this subsection, we present the decay rates including $\phi$ and $h$ which are newly studied in this paper.

One of the relevant interaction Lagrangians is the gauge interaction with $Z'$ given by
\begin{align}
\mathcal{L}_{\mathrm{gauge}} = 2 g_{B-L} m_{Z'} \cos\alpha Z'_\mu Z'^\mu \phi + 4 g_{B-L}^2 \cos^2\alpha Z'_\mu Z'^\mu \phi^2, \label{eq:lag1}
\end{align}
and the other one is the Yukawa interactions given by
\begin{align}
\mathcal{L}_{\mathrm{yukawa}} = - \frac{\sqrt{2} m_f}{v} \sin\alpha \overline{f} f \phi  
- g_{B-L}\frac{\sqrt{2} m_N}{m_{Z'}} \cos\alpha  \overline{N_i^C} N_i \phi,  \label{eq:lag2}
\end{align}
where the Yukawa coupling of a fermion $f$ is replaced by its mass $m_f$ and $v$.  The partial decay rates of $\phi \to Z'Z'$ and $NN$, $f\bar{f}$ 
are given by
\begin{subequations}
\begin{align}
\Gamma(\phi\rightarrow Z'Z') &=(g_{B-L}\cos\alpha)^2 \frac{m_\phi}{4\pi} \sqrt{1 - \frac{4 m_{Z'}^2}{m_\phi^2}} 
\frac{m_\phi^2 - 4 m_{Z'}^2+12 \frac{m_{Z'}^4}{m_\phi^2}}{m_{Z'}^2} ,\\
\Gamma(\phi\rightarrow N_iN_i) &= (g_{B-L} \cos\alpha)^2 \frac{ m_\phi}{4 \pi} \sum_i  \left(\frac{m_{N_i}}{ m_{Z'}}\right)^2  
\left(1 -\frac{4 m_{N_i}^2}{m_\phi^2} \right)^{3/2},\\
\Gamma(\phi\rightarrow f\bar{f}) &= \sin^2\alpha \frac{m_\phi}{4\pi} \left( \frac{m_f}{v}\right)^2 \left( 1 -\frac{4 m_f^2}{m_\phi^2} \right)^{3/2}.
\end{align}
\end{subequations}
The scalar boson $\phi$ also can decay into the SM Higgs bosons, $\phi \to hh$.  
However, we consider the case where $\phi$ is lighter than $h$, and therefore such production is kinematically forbidden.  
On the other hand, the decay of $h$ into $\phi \phi$ is possible and its decay rate is given by 
\begin{align}
\Gamma_h(h \rightarrow \phi \phi) &= \frac{\sqrt{m_h^2-4 m_{\phi}^2}}{16\pi m_h^2} |C_{h\phi\phi}|^2 , \label{rate:h->phiphi} 
\end{align}
with
\begin{align}
C_{h\phi\phi} &=\frac{\sin(2\alpha) (m_h^2+2 m_{\phi}^2)(m_{Z'} \sin\alpha-2 g_{B-L} v \cos\alpha)}{2 v m_{Z'}}.
\end{align}

\section{The Boltzmann equation}
\label{sec:bolt}

In this section, we show the Boltzmann equations of $N$, $Z'$ and $\phi$ for freeze-in production.
The Boltzmann equations for the number density of $N$, $Z'$ and $\phi$ are given by
\begin{subequations}
\begin{align}
 \frac{d n_{N}}{dt} + 3 H n_{N} =& \sum_{i,j=f,Z'} \langle\sigma v(ij \rightarrow NN)\rangle (n_i n_j -n_N^2)
  + \sum_{i =Z', \phi, h} \langle \Gamma(i \rightarrow NN)\rangle n_i , \label{Eq:BoltN_n} \\
 \frac{d n_{Z'}}{dt} + 3 H n_{Z'} =& \sum_{i,j=f} \langle\sigma v (ij\rightarrow Z' Z') \rangle( n_i n_j - n_{Z'}^2) 
 + \sum_{i =\phi, h} \langle\Gamma( i \rightarrow  Z'Z')\rangle  n_i  \nonumber \\
 & - \sum_{i =\phi, h} \langle\sigma v( Z'Z' \to i)\rangle  n_{Z'}^2  
   + \sum_{i,j,k = f, \gamma, G} \langle\sigma v ( Z' i\rightarrow jk)  n_i \rangle \left( n_{Z'}- n^{\mathrm{eq}}_{Z'} \right) 
 \nonumber \\
 &  - \sum_{i,j=f} \langle\Gamma(Z'\rightarrow i j)\rangle  (n_{Z'}- n^{\mathrm{eq}}_{Z'}), \label{Eq:BoltX_n} \\
  \frac{d n_{\phi}}{dt} + 3 H n_{\phi} =& \sum_{i,j=f} \langle\sigma v (ij\rightarrow \phi \phi) \rangle( n_i n_j - n_{\phi}^2) \nonumber \\
   & + \langle\sigma v( Z'Z' \to \phi)\rangle  n_{Z'}^2  - \langle\Gamma(\phi\rightarrow Z' Z')\rangle n_{\phi}  
  - \sum_{i,j=f} \langle\Gamma(\phi\rightarrow i j)\rangle  (n_{\phi}- n^{\mathrm{eq}}_{\phi}), \label{Eq:BoltP_n} 
\end{align}
 \label{eq:boltz-eqs}
\end{subequations}
where $i, j$, and $k$ are possible initial and final states in reactions, and $n_i$ is the number density of $i$-particle, respectively. 
The Hubble expansion rate in the radiation-dominated era is denoted by $H$. Note that the decay $h \to \phi \phi$ is omitted in Eq.~\eqref{Eq:BoltP_n} because its decay rate is further suppressed by $g_{B-L}^2 m_h^2/m_{Z'}^2$ than that of $\phi \to f\overline{f}$ for small $g_{B-L}$ and $\alpha$.

Thermally averaged product of the scattering cross section and relative velocity $\langle \sigma v \rangle$ in the right hand side of Eqs.~(\ref{eq:boltz-eqs}) are defined by~\cite{Gondolo:1990dk}
\begin{align}
	\langle\sigma v\rangle n_i n_j & =
	\frac{T}{32\pi^4}\sum_{i, j}\int^{\infty}_{(m_i+m_j)^2} ds g_i g_j p_{ij} 4E_{i}E_j \sigma v K_1\left(\frac{\sqrt{s}}{T}\right) ,
\end{align}
and  
\begin{align}
4E_{i}E_j \sigma v 
& \equiv \prod_f \int \frac{d^3 p_f}{(2\pi)^3}\frac{1}{2E_f}\overline{|\mathcal{M}|^2}(2\pi)^4 \delta^{(4)}(p_i+p_j-\sum p_f) \nonumber \\
 & = \frac{1}{16 \pi}\frac{2|q_f|}{\sqrt{s}}\int \overline{|\mathcal{M}|^2} d\cos\theta , \\
2|q_f| & = \sqrt{s-4 m_N^2}, \\
 p_{ij} & \equiv \frac{\sqrt{s-(m_i+m_j)^2}\sqrt{s-(m_i-m_j)^2}}{2\sqrt{s}}.
\end{align}
Here, $i$ denotes an initial state with the mass $m_i$, energy $E_i$, and internal degrees of freedom $g_i$.
The center-of-mass energy squared is given by $s = (E_i + E_j)^2$, and three-momentum of a final-state particle is denoted by $q_f$.
$K_i(z)$ is the modified Bessel function of the $i$th kind.
For the production processes of $Z'$ with a photon $\gamma$ or a gluon $G$, the thermal average of $\sigma v n$ is defined by
\begin{align}
\langle \sigma v n_i \rangle n_{Z'}^{\mathrm{eq}}
& \equiv \frac{T}{32\pi^4}\int_{(m_i+m_{Z'})^2}^{\infty} ds g_i g_{Z'} p_{i{Z'}} (4E_i E_{Z'} \sigma v)K_1\left(\frac{\sqrt{s}}{T}\right) , 
\end{align}
where $i=\gamma,G, f,\bar{f}$ and 
\begin{align}
n_{Z'}^{\mathrm{eq}} &= \frac{T}{2\pi^2}g_{Z'}m_{Z'}^2K_2\left(\frac{m_{Z'}}{T}\right).
\end{align}
Note that the gluon contribution is taken into account above $T \simeq \Lambda_{QCD} = 150$ MeV. 
Thermally averaged decay rate $\langle \Gamma \rangle$ of $i \rightarrow jj$ is defined by 
\begin{align}
	\langle\Gamma(i \rightarrow jj)\rangle =\frac{ K_1\left(\frac{m_i}{T}\right)}{K_2\left(\frac{m_i}{T}\right) } \Gamma(i \rightarrow jj).
\end{align}
The ratio of the modified Bessel function represents a suppression factor for a high temperature, $T \gg m_i$, by the time dilation. 
Concrete forms of the invariant squared amplitudes and decay rates for the processes involving only $N$ and $Z'$ are given at Appendix in Ref.~\cite{Eijima:2022dec}. 
The amplitudes for $ij \to \phi \phi$ are given in Appendix.
Note that, the thermal averaged cross sections in the first term in the right-hand-side of Eq.~\eqref{Eq:BoltN_n} should be interpreted as the one from which the $s$-channel contribution mediated by the on-shell $\phi$ is subtracted~\cite{Kolb:1979qa}. This is to avoid double-counting with the second term.
In addition, when we calculate $f\bar{f}\to Z'\gamma,~Z' G$, we introduce the thermal mass of the photon and gluon to regulate $t(u)$-channel process \cite{Redondo:2008ec}.

\section{Abundance of sterile neutrino DM} 
\label{sec:abundance}
In this section, we show our numerical results of the sterile neutrino DM abundance.
Throughout this section, the DM mass $m_N$ and the other sterile neutrino mass $m_{\nu_R}$ 
are fixed to be $m_N=0.5$ GeV and $m_{\nu_R}=2$ GeV, respectively, unless stated.
We consider the case where $Z'$ cannot decay into a DM pair, hence the $Z'$ mass is restricted to $m_{Z'} \leq 1$ GeV. 
There are two cases of mass spectrum on $m_{\phi} > 2 m_N$ and $m_{\phi} < 2 m_N$. 
We consider those in order.
Since we find $m_\phi^2 \simeq (\lambda_2)(g_{B-L}^2)^{-1} m_{Z'}^2$,
from Eqs.~\eqref{eq:Z'-mass} and \eqref{eq:phi-mass}, the mass spectrum that $m_\phi$ and $m_{Z'}$ are
of the same order, which we will study, is realized for
$\lambda_2 \sim g_{B-L}^2= 10^{-20}( g_{B-L}/10^{-10})^2$.
From Eq.~\eqref{eq:sinalpha}, we find the typical order of $\lambda_3 \sim (m_h/m_{Z'}) g_{B-L} \alpha = 10^{-15}({\rm GeV}/m_{Z'})(g_{B-L}/10^{-10})(\alpha/10^{-7})$.

\subsection{$m_{\phi} > 2m_N$ case} 
\label{subsec:phi>2N}

In this spectrum, the DM is produced from the decay $\phi \rightarrow NN$.
%
\begin{figure}[t]
\centering
\includegraphics[width=0.55\textwidth]{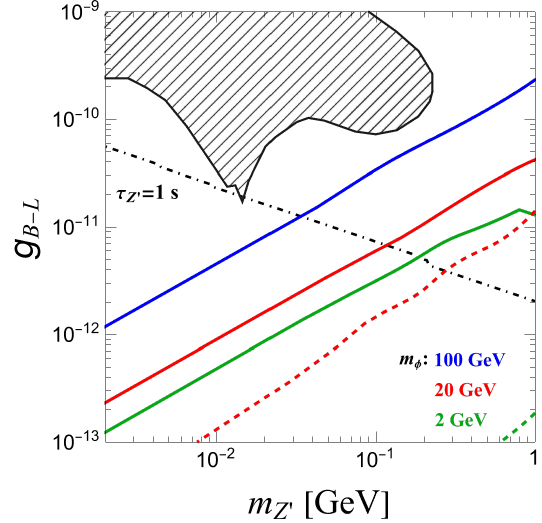}
\caption{
Contour plots for the observed DM relic abundance $\Omega h^2 = 0.12$.
The DM mass and the heavier sterile neutrino masses are fixed to 
$m_N = 0.5$ GeV and $2$ GeV, respectively.
The blue, red, and green curves correspond to $m_\phi=100$, $20$, and $2\,{\rm GeV}$, respectively.
The scalar mixing angle is taken to $\alpha = 0$ (solid) and $10^{-7}$ (dashed).
Note that in the case of $m_\phi = 100\,{\rm GeV}$ (blue), the solid and dashed lines overlap.
The dashed-dotted line corresponds to $\tau_{Z'}=1$ second.
The black shaded region is excluded by SN1987A.
 }
\label{fig:DM-abundance}
\end{figure}
%
Figure \ref{fig:DM-abundance} shows the contour plot of the DM
abundance in $m_{Z'}$--$g_{B-L}$ plane. 
The blue, red, and green curves correspond to $m_\phi=100$, $20$, and $2$ GeV, respectively. 
The scalar mixing angle is taken to be $\alpha = 0$ (solid) and $10^{-7}$ (dashed).
Note that in the case of $m_\phi = 100\,{\rm GeV}$ (blue), the solid and dashed lines overlap.
The dashed-dotted line corresponds to the lifetime of $Z'$; $\tau_{Z'} = 1$ second. 
The black shaded region is excluded by SN1987A~\cite{Croon:2020lrf}.
The gauge coupling constant to reproduce the correct DM abundance turns out to
 be much smaller than the previous study~\cite{Eijima:2022dec}. In the previous study, the dominant production modes have been regarded as the scattering $Z'Z' \rightarrow NN$, of $t(u)$-channel $N$ exchange and $s$-channel scalar $h$ and $\phi$ exchange, from thermalized $Z'$ initial states. 
However, what we found here are as follows. 
The main production mode of $\phi$ is the inverse decay $Z'Z' \to \phi$ for $\alpha = 0$ and $fG \to f \phi,~f\bar{f} \to \phi G$ 
for $\alpha = 10^{-7}$.  
In the former case, when $Z'$ is thermalized, $\phi$ can be easily produced through the inverse decay $Z'Z'\rightarrow \phi$. 
To avoid the overproduction of the DM $N$, the $\phi$ production as well as the $Z'$ production by the scattering $f G \to f Z'$ and $f\bar{f}\to Z'G $ should be suppressed. Hence, $g_{B-L}$ must be small. In the latter case, the SM fermions and gluon in the thermal bath produce $\phi$ through the scatterings. Once the scalar mixing is fixed, the production rate of $\phi$ is determined. Then, the DM abundance is controlled only by the decay of $\phi$ into $NN$. Then, $g_{B-L}$ must be smaller than the former case. 

Apart from the DM production, the late time decay of $Z'$ into the SM fermions could cause problems.  
When $Z'$ decays after the Big-Bang nucleosynthesis (BBN), the synthesized elements would be destructed by energetic particles produced by the decay and the success of BBN would be spoiled, as TeV mass gravitinos do~\cite{Khlopov:1984pf,Ellis:1984eq}.
The BBN constraints on sub-GeV decaying particles have been derived in Ref.~\cite{Kawasaki:2020qxm}.
For reference purpose, we add the dashed-dotted line corresponding to the $Z'$ lifetime of $1$ second in Fig.~\ref{fig:DM-abundance}: below the line the $Z'$ boson would decay after the BBN starts. 

\begin{figure}[htbp]
\begin{center}
\includegraphics[width=0.47\textwidth,clip]{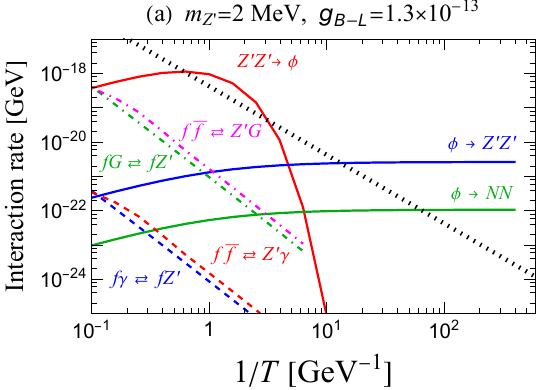}~~
\includegraphics[width=0.47\textwidth,clip]{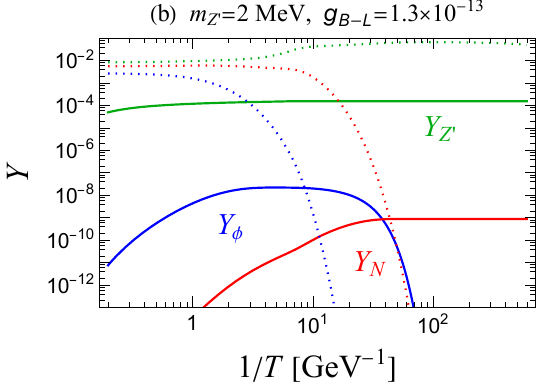}  \vspace{4mm} \\ 
\includegraphics[width=0.47\textwidth,clip]{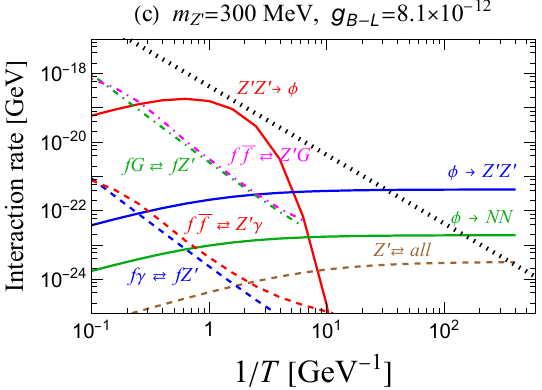}~~ 
\includegraphics[width=0.47\textwidth,clip]{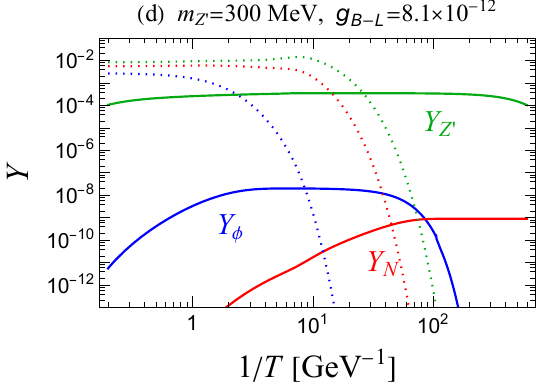}\vspace{4mm} \\ 
\includegraphics[width=0.47\textwidth,clip]{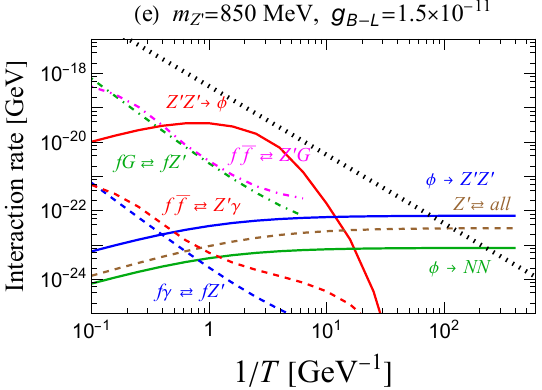}~~
\includegraphics[width=0.47\textwidth,clip]{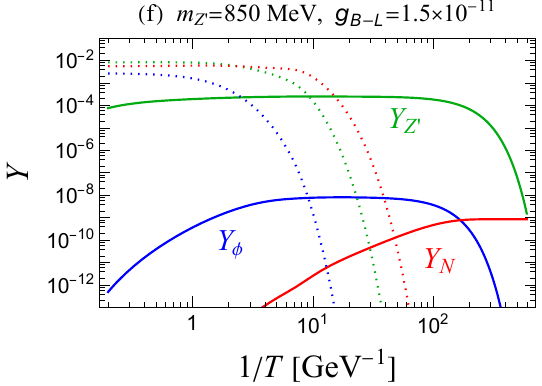}
\end{center}
\vspace{-5mm}
\caption{Time evolution of interaction rates (left) and yield (right).
In all panels, the mass of $\phi$ and $N$, and scalar mixing angle are fixed as $m_\phi = 2$ GeV, $m_N^{}=0.5$ GeV, and $\alpha = 0$, respectively.
In the left panels, the black dotted line represents the Hubble parameter in the radiation dominated Universe.
In the right panels, the dotted lines represent the time evolution assuming that each particle is in thermal equilibrium.
}
\label{fig:mphi2GeV-alpha0}
\end{figure}
\begin{figure}[htbp]
\begin{center}
\includegraphics[width=0.47\textwidth,clip]{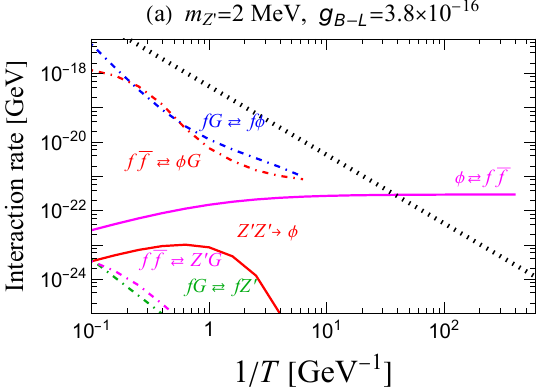}~~
\includegraphics[width=0.47\textwidth,clip]{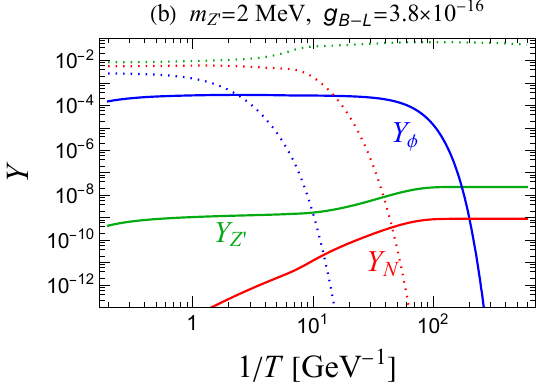}  \vspace{4mm} \\ 
\includegraphics[width=0.47\textwidth,clip]{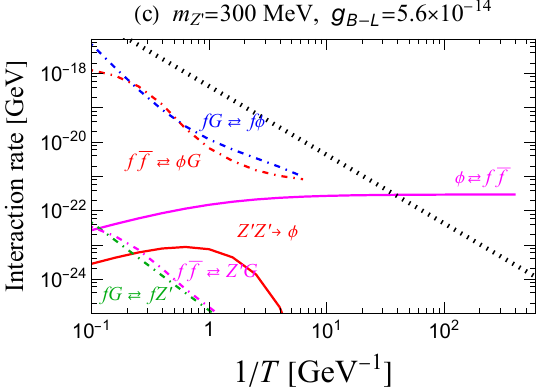}~~ 
\includegraphics[width=0.47\textwidth,clip]{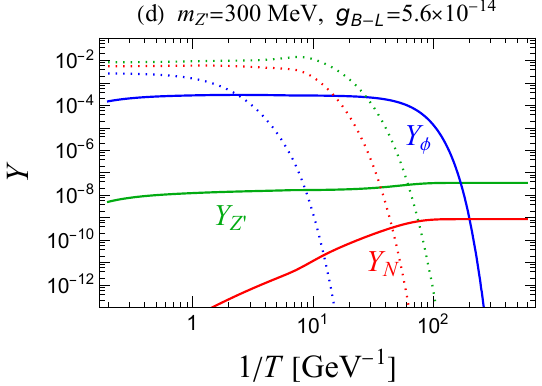}\vspace{4mm} \\ 
\includegraphics[width=0.47\textwidth,clip]{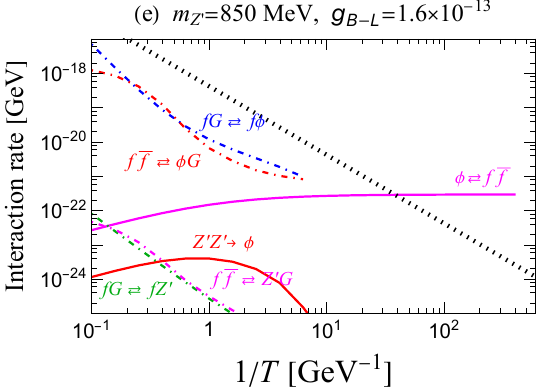}~~
\includegraphics[width=0.47\textwidth,clip]{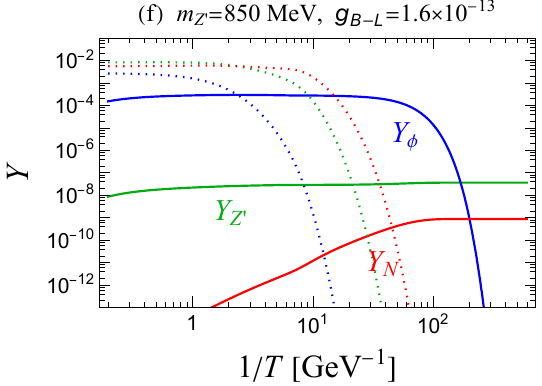}
\end{center}
\caption{
Time evolution of interaction rates (left) and yield (right) in the case of $\alpha = 10^{-7}$.
The mass of $\phi$ and $N$ are fixed as $m_\phi = 2$ GeV and $m_N^{}=0.5$ GeV, respectively.
 }
\label{fig:mphi2GeV-alpha10-7}
\end{figure}

Figures \ref{fig:mphi2GeV-alpha0} and \ref{fig:mphi2GeV-alpha10-7} show the time evolution of 
interaction rate $\langle\sigma v (ij \rightarrow kl) \rangle s$ and  $\langle\Gamma (i\rightarrow jk) \rangle$ (panel (a), (c) and (e)), 
and the yield value $Y$ of $N$, $Z'$ and $\phi$ ((b), (d) and (f)) for benchmark points that satisfy $\Omega h^2=0.12$.
Here, $s$ is the entropy density. The black dotted line represents the Hubble parameter in the radiation dominated Universe, $H \simeq T^2/M_{pl}$ with the Planck mass $M_{pl} = 2.4 \times 10^{18}$ GeV. 
In Fig.~\ref{fig:mphi2GeV-alpha0}, the scalar mass and mixing angle
are fixed to be $m_\phi = 2$ GeV and $\alpha=0$, respectively.
The values of $m_{Z'}$ and $g_{B-L}$ are varied in each row: 
the first, second, and third row of figures correspond to the case 
$(m_{Z'},g_{B-L})=(2~{\rm MeV}, 1.3\times 10^{-13})$, 
$(300~{\rm MeV}, 8.1\times 10^{-12})$, 
and $(850~{\rm MeV}, 1.5\times 10^{-11})$, respectively.
In the left panels, solid curves represent the rates of interactions involving $\phi$, dash-dotted ones represent those involving gluon, while the other interaction rates are plotted with dashed lines.
One can see from the left panels that two new processes studied in this paper are dominant production modes of $\phi$ and $Z'$.
One is the inverse decay $Z'Z' \to \phi$ which is much larger than $f\bar{f} \to \phi$. 
From the inverse decay, $\phi$ is efficiently produced and then decays into the dark matters roughly below $T \sim 0.1$ GeV from the $\phi$ decay.
The second one is the fermion-gluon scattering $f G \to f Z'$ and $f\bar{f}\to Z'G$ which can be large in gluon bath and efficiently produces $Z'$.
In the right panels in Figs.~\ref{fig:mphi2GeV-alpha0} and \ref{fig:mphi2GeV-alpha10-7}, solid curves represent the yield of $Z'$ (green), $\phi$ (blue) and $N$ (red) obtained in our calculation while dashed ones are equilibrium values. 
One can see that $\phi$, $Z'$ and $N$ never enter the thermal bath and are produced in freeze-in mechanism.
As seen in Fig.~\ref{fig:mphi2GeV-alpha0}-(b), (d), and (f), the temperature at which $Y_\phi$ begins to decrease becomes lower as the $m_{Z'}$ becomes larger. 
This can be understood as follows. 
$Y_{\phi}$ starts to decrease when the $\phi\to Z'Z'$ process becomes comparable to the Hubble parameter. 
The larger $m_{Z'}$ leads to the smaller decay rate of $\phi\to Z'Z'$.

Figure \ref{fig:mphi2GeV-alpha10-7} is the same plot as Fig.~\ref{fig:mphi2GeV-alpha0} 
for $\alpha = 10^{-7}$. 
The scalar mass is fixed to be $m_\phi = 2$ GeV.
The values of $m_{Z'}$ and $g_{B-L}$ are varied in each row: 
the first, second, and third row of figures correspond to the case 
$(m_{Z'},g_{B-L})=(2~{\rm MeV}, 3.8\times 10^{-16})$,
$(300~{\rm MeV}, 5.6\times 10^{-14})$, 
and $(850~{\rm MeV}, 1.6\times 10^{-13})$, respectively.
As seen in Fig.~\ref{fig:mphi2GeV-alpha10-7}-(b), (d), and (f), the evolution of $Y_{Z'}$ and $Y_\phi$ are completely different from those in Fig.~\ref{fig:mphi2GeV-alpha0}.
This is due to the fact that $\phi$ can interact with the SM particles through scalar mixing. 
The left panels show that $\phi$ is produced through not only $Z'Z' \to \phi$ but also 
$f\bar{f}\to \phi$, $fG\to f\phi$, and $f\bar{f}\to \phi G$.
That is why the maximal values of $Y_\phi$ in Fig.~\ref{fig:mphi2GeV-alpha10-7} become larger than 
those in Fig.~\ref{fig:mphi2GeV-alpha0}.
Contrary to Fig.~\ref{fig:mphi2GeV-alpha0}, the temperature at which 
$Y_\phi$ begins to decrease is common in Fig.~\ref{fig:mphi2GeV-alpha10-7}-(b), (d), and (f).
This is because $Y_{\phi}$ decrease through $\phi \to f\bar{f}$, $f\phi\to fG$, and $\phi G\to f\bar{f}$,
all of which are independent of the gauge coupling $g_{B-L}$.

\subsection{$m_{\phi} < 2m_N$ case} 
\label{subsec:phi<2N}

\begin{figure}[htbp]
\centering
\includegraphics[width=0.55\textwidth]{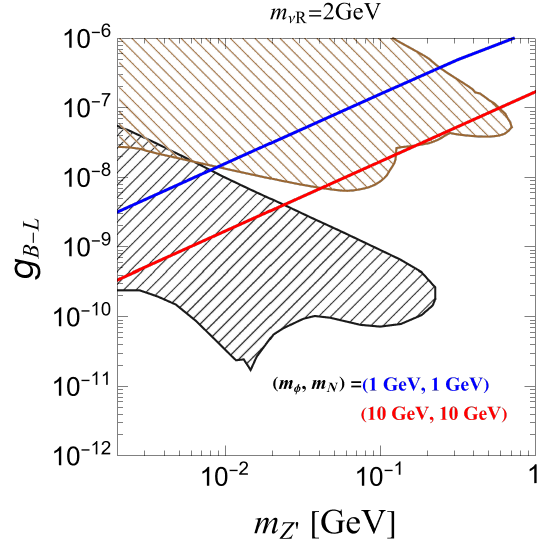}
\caption{
Blue and red contour plots of $\Omega h^2 = 0.12$ for $(m_\phi, m_N) = (1$ GeV$, 1$ GeV$)$ and $(10$ GeV$,10$ GeV$)$, respectively. 
The other sterile neutrino mass $m_{\nu_R}$ is taken to be $2$ GeV.
The black shaded region is excluded by SN1987A.
The brown shaded region is excluded by electron and proton beam dump experiments.
 }
\label{fig:DM-abundance_faser}
\end{figure}
Figure \ref{fig:DM-abundance_faser} is the same plot as Fig.~\ref{fig:DM-abundance} for the mass spectrum 
$(m_\phi, m_N) = (1$ GeV$, 1$ GeV$)$ (blue) and $(10$ GeV$,10$ GeV$)$ (red).
The brown shaded region is excluded by electron and proton beam dump experiments  
(See Refs.~\cite{Feng:2022inv,Asai:2022zxw} and references therein).
For this mass spectrum, $\phi$ cannot decay into $N$ and therefore the DM must be produced through the scattering of $Z'Z' \to NN$, dominantly $t(u)$-channel $N$ exchange.
Hence the results is insensitive to the scalar mixing $\alpha$ and essentially same as the case B in Ref.~\cite{Eijima:2022dec}. In this case, the $Z'$ boson search will be possible at FASER2~\cite{Feng:2022inv, Araki:2020wkq, Asai:2022zxw} and SHiP~\cite{Alekhin:2015byh} experiment.

\section{Conclusion}
\label{sec:concl}
We have studied freeze-in production of sterile neutrino dark matter in feebly interacting $B-L$ model. In the spectrum where $Z'$ cannot decay 
into the dark matter, the scalar plays an important role in the dark matter production. 
We solved the Boltzmann equations of $Z',~N$ and also $\phi$ simultaneously and analyzed the scalar contribution to the dark matter production.
We also took the $Z'$ production with a gluon into account.

We found that $Z'$ can be predominantly produced by $fG \to f Z'$, $f\bar{f}\to Z'G$ and produces $\phi$ through the inverse decay of $Z'Z' \to \phi$ when we turn off the scalar mixing $\alpha$. 
Then, the sterile neutrino DM is produced by the decay of $\phi$.  
We showed that not only the DM but also $Z'$ and $\phi$ never enter the thermal bath and are produced in freeze-in mechanism.
When the scalar mixing $\alpha$ is nonzero, $\phi$ also is additionally produced by the inverse decays of the SM particles, $fG\to f\phi$, and $f\bar{f}\to \phi G$.
For very small $g_{B-L}$ coupling, $Z'$ may be long-lived and subject of the constraints from the BBN.

We also found when the $\phi$ decay into the DM is forbidden, the gauge coupling should be $10^{-10}$--$10^{-6}$ for $m_{Z'} \leq 1$ GeV. Such a region will be searched by collider experiments like FASER2~\cite{FASER:2018eoc,FASER:2023tle} and SHiP~\cite{Alekhin:2015byh}. We will show its sensitivity region in our future work. 

\section*{Acknowledgments}
We thank Shintaro Eijima for valuable discussion on the early state of this work. We also thank Masaharu Tanabashi for important comments.
This work is supported, in part, by JSPS KAKENHI Grants No.~JP19K03865, No.~JP23K03402 (O.S.), 
and JSPS KAKENHI Grants No.~22K03622 and 23H01189 (T.S.),
 by National Natural Science Foundation of China Grant No.~NSFC-12347112 (Y.U.), 
 and by the Guangdong Major Project of Basic and Applied Basic Research No.~2020B0301030008 (Y.U.).

\appendix

\section{Amplitude of the scalar production processes}

%
We give explicit formulas of the amplitude. The scalar self-couplings are given in  \cite{Eijima:2022dec}.

\subsection{$ f(p_1)\bar{f}(p_2) \rightarrow \phi(q_1) \phi(q_2)$}

%
\begin{align}
M = M_s + M_{tu},
\end{align}
where
\begin{subequations}
\begin{align}
M_s &= \frac{\sqrt{2} m_f}{v} \overline{v}(p_2) u(p_1) 
\left( 
\frac{C_{\phi\phi\phi} \sin\alpha}{(p_1+p_2)^2 - m_\phi^2 + i m_\phi \Gamma_\phi}
+\frac{C_{h\phi\phi} \cos\alpha}{(p_1+p_2)^2 - m_\phi^2 + i m_\phi \Gamma_\phi}
\right), \\
M_{tu} &= -2 \frac{m_f^2}{v^2} \sin^2\alpha \nonumber \\
&\quad \times \overline{v}(p_2)\left(
\frac{(\slashed{p_1}-\slashed{q_1}) + m_f}{(p_1-q_1)^2 - m_f^2 + im_f \Gamma_f}
+\frac{(\slashed{p_1}-\slashed{q_2}) + m_f}{(p_1-q_2)^2 - m_f^2 + im_f \Gamma_f}
\right) u(p_1),
\end{align}
\end{subequations}
and
\begin{align}
C_{\phi\phi\phi} &=3 \frac{ m_{\phi }^2 \left( 2 g_{B-L} v \cos^3\alpha +m_{Z'} \sin^3\alpha \right)}{v m_{Z'}}.
\end{align}

\subsection{$ Z'(p_1)Z'(p_2) \rightarrow \phi(q_1) \phi(q_2)$}

\begin{align}
M = M_c + M_s + M_{tu},
\end{align}
where
\begin{subequations}
\begin{align}
M_c &= 16 g_{B-L} \cos^2\alpha \epsilon^\mu(p_1) \epsilon_\mu(p_2), \\
M_s &= -g_{B-L} m_{Z'} \epsilon^\mu(p_1) \epsilon_\mu(p_2) \nonumber \\
&\quad \times \left( 
\frac{C_{\phi\phi\phi} \cos\alpha}{(p_1+p_2)^2 - m_\phi^2 + i m_\phi \Gamma_\phi}
-\frac{C_{h\phi\phi} \sin\alpha}{(p_1+p_2)^2 - m_\phi^2 + i m_\phi \Gamma_\phi}
\right), \\
M_{tu} &= 4 g_{B-L}^2 m_{Z'}^2 \cos^2\alpha \epsilon^\mu(p_1) \epsilon_\mu(p_2) \nonumber \\
&\quad \times \left(
\frac{1}{(p_1-q_1)^2 - m_{Z'}^2 + im_{Z'} \Gamma_{Z'}}
+\frac{1}{(p_1-q_2)^2 - m_{Z'}^2 + im_{Z'} \Gamma_{Z'}}
\right).
\end{align}
\end{subequations}

\subsection{$ \gamma(q_1)f(q_2) \rightarrow \phi(p_1)f(p_2) $}

%
\begin{align}
M = M_s + M_t,
\end{align}
where
\begin{subequations}
\begin{align}
M_s &= \frac{\sqrt{2} e m_f q_f\sin\alpha \overline{u}(p_2)\left( (\slashed{q_1}+\slashed{q_2})+m_f\right)\gamma^{\mu}u(q_2) \epsilon_\mu(q_1)}{ v \left( (q_1+q_2 )^2-m_f^2 \right)}, \\
M_{t} &= \frac{\sqrt{2} e m_f q_f\sin\alpha \overline{u}(p_2)\gamma^{\mu}\left( (\slashed{q_2}-\slashed{p_1})+m_f\right)u(q_2)  \epsilon_\mu(q_1)}{v \left( (q_2-p_1)^2-m_f^2\right)}.
\end{align}
\end{subequations}
with $e$ being the electromagnetic coupling with its charge $q_f$ of a fermion $f$.

For gluon $G$, we need to replace the coupling constant with the strong coupling $g_s$
and multiply the color factors instead of $q_f$.

\subsection{$ f(q_1)\bar{f}(q_2) \rightarrow \phi(p_1) \gamma(p_2)$}
%
\begin{align}
M = M_t + M_u,
\end{align}
where
\begin{subequations}
\begin{align}
M_s &= \frac{\sqrt{2} e m_fq_f\sin\alpha \overline{v}(q_1)\gamma^{\mu}\left( (\slashed{q_2}-\slashed{p_1})+m_f\right)u(q_2) \epsilon^\ast_\mu(p_2)}{v \left( (q_2-p_1)^2-m_f^2 \right)}, \\
M_{t} &= \frac{\sqrt{2} e m_fq_f\sin\alpha \overline{v}(q_1)\left( (\slashed{q_2}-\slashed{p_2})+m_f\right)\gamma^{\mu}u(q_2) \epsilon^\ast_\mu(p_2)}{v \left( (q_2-p_2)^2-m_f^2\right)}.
\end{align}
\end{subequations}

\bibliography{biblio}

\end{document}